\documentclass[preprint,eqsecnum,aps]{revtex4}

\usepackage{graphicx}
\usepackage{dcolumn}
\usepackage{amsmath}

\makeatletter
\begin{document}
%\baselineskip 9mm \pagestyle{empty}
%\preprint{ANU-TH-03-}

\title{Elliptic Flow and Jet Quenching of a Parton System after Relativistic Heavy Ion Collision}

\author{Ghi R. Shin}
\affiliation{Department of Physics, Andong National University,
                    Andong, South Korea}

\date{\today}
~ \\

\begin{abstract}
We obtain the initial phase space distribution after relativistic heavy ion collision
by the CGC shattering method incorporating the uncertainty principle
and solve the semi-classical Boltzmann equation which includes the gluon radiation processes.
We present as a function of time the attenuation rate of high $p_T$ partons, 
which have transverse momenta over 6 $GeV/c$, in the medium which is formed
after relativistic heavy ion collision. We calculate the elliptic flow as a function of 
an impact parameter, time and transverse momentum and also present the polar anisotropy, 
which gives the initial condition for color filamentation.
\end{abstract}

\maketitle
\newpage
\section{Introduction}
One of conclusions drown from the experiments at RHIC for the first period of experiment
is that the strongly interacting Quark-Gluon Plasma is formed after Au-Au collision at
maximum energy\cite{star05,phen05}. 
This conclusion, which is widely accepted by the community, is compelling
based on several experimental results: The elliptic flow data, which is understood to be
buildup at early stage of collision process, can be explained by hydrodynamical evolution
of a plasma. And away-side jet quenching can be understood by the energy loss into 
the dense medium.

It however is worth to reconsider the results in parton cascade simulation whether the data 
can be reproduced with no hydrodynamics. This is the purpose of our note.

We describe briefly the Monte-Carlo simulation solving the semi-classical  Boltzmann equations
of motion in the next section.
We obtain the initial phase-space distribution of partons in Section III. 
We presenr the results of our numerical solutions, namely, elliptic flow and 
polar anistropy in Section IV and the jet quenching in Section V.
We summarize and conclude in Section VI.

\section{Parton Cascade Code}
Although a system of partons evolves according to the field equation of quantum field theory,
QCD, it is much more intuitive and rich in structure to use quantum transport equations\cite{elze,geiger} 
based on the field theory. It however is very difficult to solve the equations and 
interprete the functions which are defined on the spin and color index\cite{birula}.
The semi-classical Boltzmann equation is used thus to capture the main effects\cite{bla87}.
To solve the partonic Boltzmann equation, the partonic cascade codes\cite{gei97,gyu97} have been developed.
In the recent studies\cite{shin02,shin03}, the parton cascade code(PCC), 
which includes $gg \rightarrow ggg$ channel in the secondary collision,
has been developed and applied to the heavy ion collision. 
In the study, the minimum transverse momentum to produce
minijets from the primary collision of two nuclei is set at 1 GeV/c and the small angle scatterings,
$sin \theta \geq p_t / E$ where $p_t$ is the minimum momentum transfer set at 0.5 GeV/c, are
included.
Solving the Boltzmann transport equation with a given initial distribution
by using a Monte Carlo simulation, the author presented the energy density, number density
as well as momentum isotropicity of partons.
The detailed description to solve the equations can be found in \cite{shin02,shin03}.

We are obliged to mension a few flaw we have in solving the evolution equations:
It has been known that the parton system can be very dense in early stage after relativistic
heavy ion collision so that the partonic wave functions can overlap among them. 
This could invalidate the kinetic theory. 
On top of that we assume the particle is born just at the collision point 
with no formation time so that they can make a collsion just after collision.

The processes we consider in our study are 
\begin{eqnarray}
gg &\leftrightarrow& gg, q \bar q,\\ 
g q &\leftrightarrow& gq,\\ 
g \bar q &\leftrightarrow& g \bar q, \\
q^a q^b &\leftrightarrow& q^c q^d , \\
q\bar q &\leftrightarrow& q \bar q,\\
\bar q^a \bar q^b &\leftrightarrow& \bar q^c \bar q^d, \\
gg &\rightarrow& ggg .
\end{eqnarray}
We however miss some of basic channels such as $qg \rightarrow q \gamma$, 
$q \bar q \rightarrow \gamma \gamma$, $q \bar q \rightarrow g \gamma$,
which could provide important information on the system.

The cross sections for the processes up to the leading order (LO) are rewritten here
for convenience\cite{peskin};
\begin{eqnarray}
{{d\sigma^{gg\rightarrow gg}}\over{dt}} &=&
{{9\pi\alpha_s^2}\over{2s^2}}(3-{{tu}\over s^2} - {{su}\over{t^2}}
- {{st}\over{u^2}} ) \\ \\			\label{gg_gg}
{{d\sigma^{gg\rightarrow q_a\bar q_b}}\over{dt}} &=&
{{\pi\alpha_s^2}\over{6s^2}} \delta_{ab} ( {u \over t} + {t \over
u} - {9\over 4}{{t^2+u^2}\over{s^2}} ) \\ \\	\label{gg_qbq}
{{d\sigma^{gq\rightarrow gq}}\over{dt}} &=&
{{4\pi\alpha_s^2}\over{9s^2}} ( - {u\over s} - {s\over u} + {9
\over 4}{{s^2+u^2}\over{t^2}}) \\  \\		\label{gq_gq}
{{d\sigma^{q_aq_b\rightarrow q_aq_b}}\over{dt}} &=&
{{4\pi\alpha_s^2}\over{9s^2}}[{{s^2+u^2}\over{t^2}}
+\delta_{ab}({{t^2+s^2}\over{u^2}}-{2\over 3}{{s^2}\over{ut}})]
\\  \\		\label{qq_qq}
{{d\sigma^{q_a\bar q_b\rightarrow q_c\bar q_d}}\over{dt}} &=&
{{4\pi\alpha_s^2}\over{9s^2}}[ \delta_{ac}\delta_{bd}
{{s^2+u^2}\over{t^2}} + \delta_{ab}\delta_{cd}
{{t^2+u^2}\over{s^2}}-\delta_{abcd}{2\over 3}{u^2\over{st}}] 
\\ \\		\label{qbq_qbq}
{{d\sigma^{q_a\bar q_b\rightarrow gg}}\over{dt}} &=&
{{32\pi\alpha_s^2}\over{27s^2}} \delta_{ab} [{u \over t}+{t\over
u}-{9\over 4}{{t^2+u^2}\over{s^2}}]
\label{qbq_gg}
\end{eqnarray}
and
\begin{eqnarray}
{{d\sigma^{g\bar q\rightarrow g\bar
q}}\over{dt}} &=& {{d\sigma^{gq\rightarrow gq}}\over{dt}}, \\ 
{{d\sigma^{\bar q \bar q\rightarrow \bar q \bar q}}\over{dt}} &=&
{{d\sigma^{qq\rightarrow qq}}\over{dt}}.
\end{eqnarray}
where $s, t, u$ are Mandelstam variables.

We use the following approximation for the cross section of $gg
\rightarrow ggg$ process\cite{bir93},
%\begin{widetext}
\begin{eqnarray}
{{d\sigma^{gg\rightarrow ggg}}\over{dq_\perp^2 dy dk_\perp^2}} &=&
{{9 C_A\alpha_s^3}\over{2}}
{{q_\perp^2}\over{(q_\perp^2+\mu_D^2)^2}} \cdot\nonumber \\
&& {{\Theta(k_\perp \lambda_f - \cosh y) \Theta({{\sqrt{s}}} -
k_\perp \cosh y)}\over{k_\perp^2\sqrt{ (k_\perp^2 + q_\perp^2 +
\mu_D^2 )^2 - 4 k_\perp^2 q_\perp^2}}} 
\label{gg_ggg}
\end{eqnarray}

We set the coupling constant to be 0.3 in our program.
The minimum momentum transfer for the collision is 500 $MeV/c$ and the total cross section
for $gg \rightarrow gg$ with this momentum cutoff, for example, is about $10/GeV^2$ 
which is about 4 $mb$. We usually set the K-factor to 2 or 4 to include the higher-order diagrams.

Once the initial phase space is given we sequentially make a collision among the partons.
To find the collision event relativistically, we only use the Coulomb-like part of QCD interaction
with the assumption that the closest point between two partons is the collision event.

\section{Initial phase space distribution}
The parton phase space distribution produced after relativistic heavy ion collision is one of
most important indredients for the parton cascade simulation but also of most difficult
subjets which is still in progress to its understanding.
One of ways to obtain the distribution is to make elastic scatterings between the partons of 
projectile hadron or nuclei and those of target. 
This, so-called, factorization method, has been used to calculate jets or minijets\cite{eks96,nayak,ham99,coo02}.
There are many schemes to obtain the parton distribution of hadron or nuclei\cite{szc} which is needed for
the factorization method, for examples, we can fold the parton distribution of nucleon and the nucleon distribution
of nucleus\cite{grv98,eks99}.
This method only gives relatively high $p_T$ partons and soft partons can not be obtained so that
it gives a reasonable distribution at LHC energy but does not give good one at RHIC energy.

A quite different way to obtain the initial parton information has been proposed by Krasnitz et al\cite{knv03,lap03}
based on CGC\cite{mcl94,am00}, namely, shattering of the color glass condensates of two colliding nuclei, which
we will use in our study.

The transverse momentum distribution of the produced partons after the shattering of two
CGC is implicitly given by the eq. (24) of reference \cite{knv03}. We can choose 
the transverse momentum using the Monte Carlo sampling.
The rapidity distribution is in general flat at central rapidity region and fall off almost linearly
to zero. We assume that the distribution is flat in rapidity between -2.5 and +2.5 and falls off to zero
at $\pm 5$. The number of parton produced can be obtained by integration over transverse
momentum and rapidity. This number however depends on the saturation momentum, $Q_s$,
which is introduced in CGC scheme and estimated to be $1 GeV < Q_s < 2 GeV$ at RHIC energy.
We set $Q_s = 1.7 GeV$ and the total number of initial partons is about 5400 for the
head-on collision at RHIC energy. 
Since this gives only momentum distribution, i.e., transverse momentum and rapidity,
we need to make a further assumption: 
The transversal position of produced parton can be choosen any point within 
the transverse overlap region of two colliding nuclei assuming the CGC is flat. 
On the other hand the longitudinal position will be chosen
according to the uncentainty relation in between $-1/ \Delta p_z $ and $+1/\Delta p_z$.
And the time when the partons are born will be explained in detail later.

Table 1 shows the number of partons produced after a relativistic heavy ion collision
when the saturation momentum of color glass condensate is chosen to be 1.7 $GeV/c$.

\begin{table}
\begin{tabular} {|c|c|} \hline
\; {$b$(fm)}\;  & \; \# of partons \;  \\ \hline
0 & 5400 \\ \hline
1 & 5180 \\ \hline
3 & 4150 \\ \hline
5 & 2800 \\ \hline
7 & 1570 \\ \hline
9 & 630 \\ \hline
\end{tabular}
\caption{The number of parton produced after relativistic heavy ion collisions.}
\end{table}

When the impact parameter is nonzero, we can calculate the number of partons and the transverse
position by assuming the CGC plate is circular of the nucleus's radius so that 
the total number of partons produced from non-central heavy ion collision is proportional to
the overlap area relative to that of head-on collision. And
we can choose the (transversal) position according to the overlap function, Eq. 3.1,
where the probability density is given by
\begin{eqnarray}
P(\vec r; \vec b) &=& T_{AB} (\vec r ;\vec b ) \\ \nonumber
&=& 4 \rho_0^A \rho_0^B \sqrt{R_A^2-(\vec r - \vec b/2)^2}
\sqrt{R_B^2-(\vec r + \vec b/2)^2},
\label{overlap_fn}
\end{eqnarray}
Sampling the position according to this probability density can be performed
using acceptance-rejectance method, namely we sample $(x,y)$ within the
allowed region randomly and calculate the probability density at that position
to give $P(x,y;\vec b)$. We generate a random number $r$ and compare
to $P(x,y;\vec b)/P(0,0;\vec b)$. If $r$ is less than $P(x,y;\vec b)/P(0,0;\vec b)$,
we accept the $(x,y)$ but if $r$ is greater than that, we reject that and sample another
positions. We next assume that those partons are born within the logitudinal uncertainty,
$\Delta z > 1/ \Delta p_z $, so that the z can be choosen between $-\Delta z$ and
$+\Delta z$. On the other hand, the initial parton's formation time will be given
by $1/ E$ even after the CGC shattering.
The average energy of partons in rapidity $|y| < 1 $ is about 1.03 GeV and that in $1< |y| < 2$
is about $2.14 GeV$, and those of $2< |y| < 3$ is $5.53 GeV$. Thus the partons of low energy has
formation time $0.1 - 0.2 fm/c$ while the high energy partons have much shorter formation
time.

\begin{figure}
\includegraphics{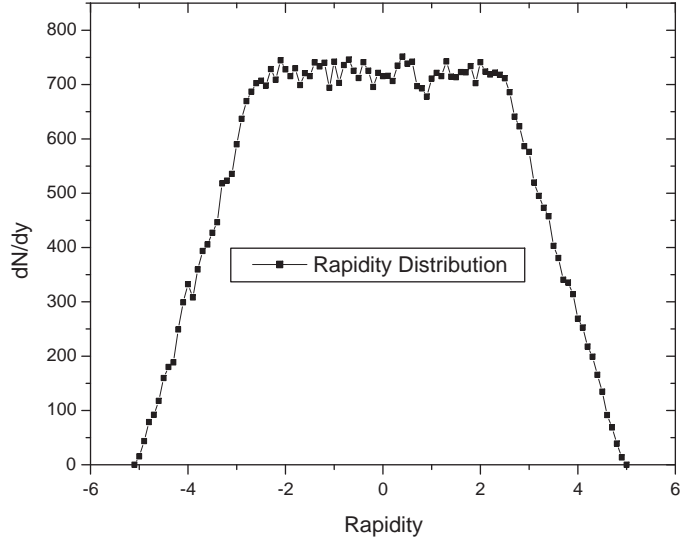}
\caption{Rapidity distribution of the test partons sampled
by a Monte Carlo method.} \label{fig1}
\end{figure}

Fig. \ref{fig1} shows the rapidity distribution of the initial state sampled by Monte Carlo
Method on head-on collision at RHIC energy. 
\\
Since the distribution is not perfectly symmetric nor flat, we expect the initial elliptic flow,
$v_2$, is not zero and fluctuate from set to set. This fluctuation shows up further analysis of jet quenching and elliptic flow.
Note that we only sample 10 sets of initial distribution because of the limited CPU time.
The total energy liberated from the CGC shattering is 34.5 TeV out of about 40 TeV of total available energy.

\begin{figure}
\includegraphics{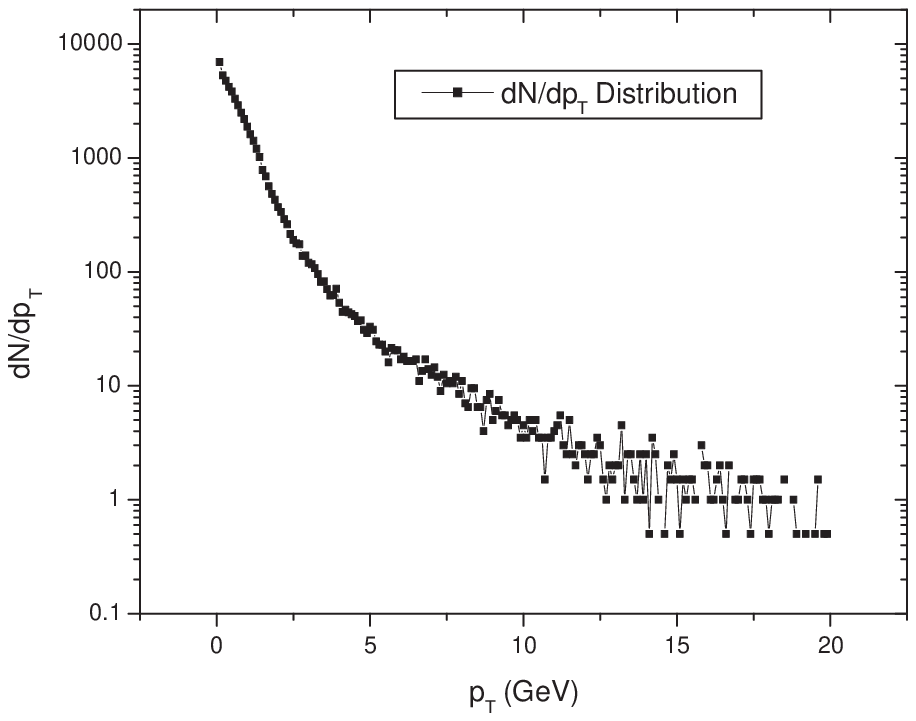}
\caption{$p_T$ distribution of the test partons sampled
by a Monte Carlo method.} \label{fig2}
\end{figure}

Fig. \ref{fig2} shows the transverse momentum distribution which can be compared with Eq.24 of ref. \cite{knv03}.
\\
The number of jets, which have the transverse momentum over 6 GeV in this study, for example,
is about 45 for Au-Au head-on collision at RHIC energy when we use the shattering mechanism
while the number reduces to about 10 for the factorization method.

\begin{figure}
\includegraphics{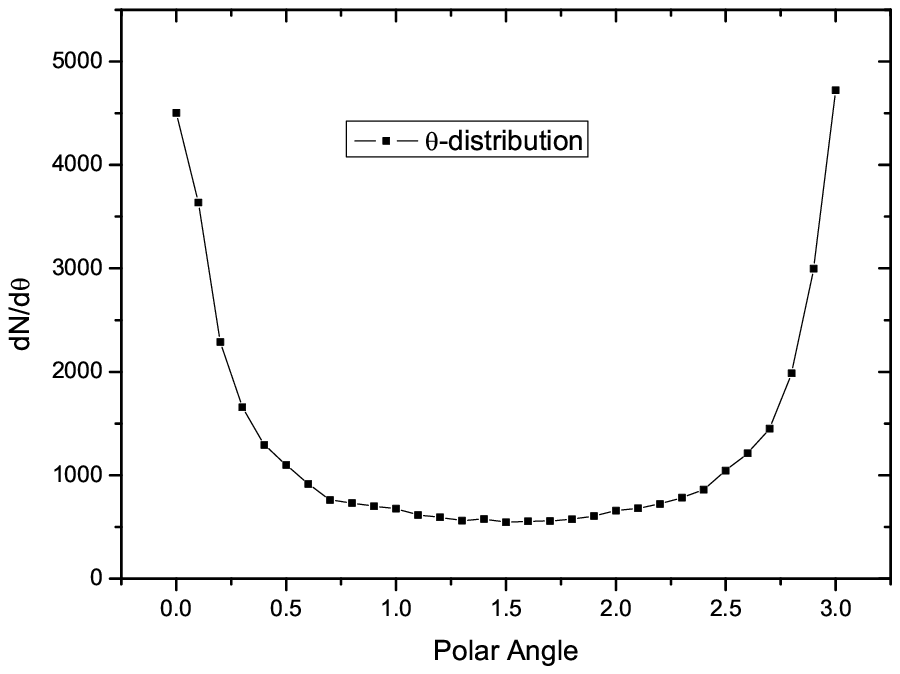}
\caption{$\theta$ distribution of the test parton's momenta sampled
by a Monte Carlo method.} \label{fig4}
\end{figure}

%Fig. \ref{fig3} shows the azimuthal angle distribution of momentum.
%\begin{figure}
%\includegraphics{dndphi.eps}
%\caption{$\phi$ distribution of the test parton's momenta sampled
%by a Monte Carlo method.} \label{fig3}
%\end{figure}\\
Fig. \ref{fig4} shows the polar angle distribution of parton's momentum and 
shows the parton are strongly forward and backword oriented 
but relatively flat in central(transversal) direction.
\\

\section{Momentum Anisotropy}
The primary partons, which are produced directly from the colliding nuclei, have azimuthal symmetry
in momentum distribution since the collision cross section between partons has azimuthal independent.
The partons will have numerous collisions among themselves after they are born and 
could develop anisotropy in momentum distribution if there is spacial anisotropic.
This azimuthal anisotropy can be extracted systematically by expanding the number density 
as a function of azimuthal angle with respect to the reaction plane\cite{oll,zgk99},
\begin{eqnarray}
E {{d^3 N} \over {dp^3}} &=& {1 \over \pi} {{d^2 N} \over {dp_T^2 dy}} [ v_0 + 2 v_1 cos \phi
+ 2 v_2 cos 2 \phi + ... ],
\end{eqnarray}
where $v_1$ is the directed flow and $v_2$ the elliptic flow. When a parton momentum is known, the elliptic
flow can be calculated by the equation,
\begin{eqnarray}
v_2 &=& < {{p_x ^2 - p_y ^2 } \over { p_x ^2 + p_y ^2 }} >
\end{eqnarray}
where the bracket denotes the average over the partons and the impact parameter vector and collision
axis defines the reaction plane. 
\begin{figure}
\includegraphics{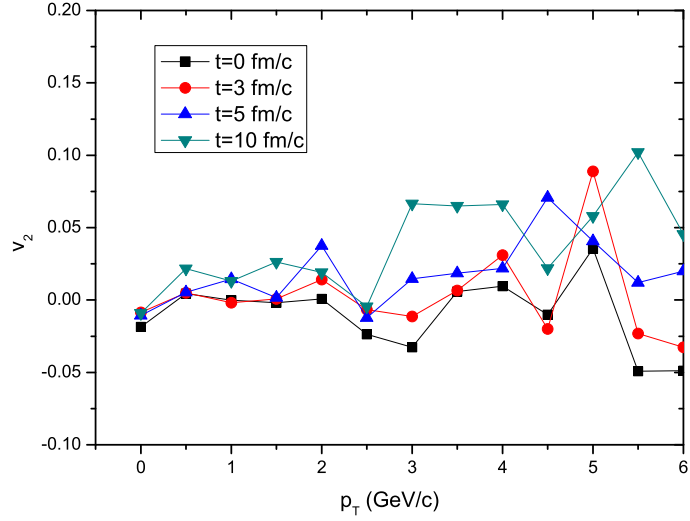}
\caption{Elliptic flow for $Q_s=1.7$ GeV and $K=4$ at $b=5$ fm.} \label{fig5}
\end{figure}\\
Fig. \ref{fig5} shows the elliptic flow as a function of transverse momentum
for the parton system formed with $Q_s = 1.7$ GeV, $K=4$ and $b=5$ fm.
We include only those of partons with $|y| < 2 $.
Since we average over 10 different sets of test partons, Fig. \ref{fig5} shows large fluctuation.
We however can see that the overall number is close to those of elliptic flow per parton shown
in Fig. 20 of referance \cite{star05}.
 
\begin{figure}
\includegraphics{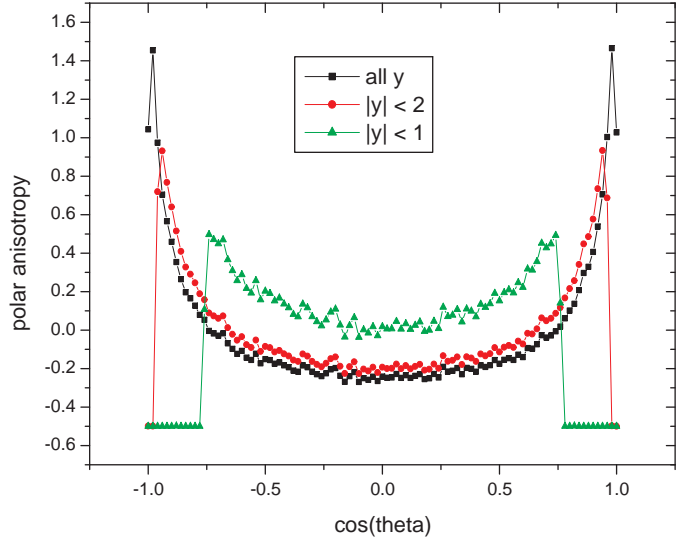}
\caption{Polar anisotropy for $Q_s=1.7$ GeV and $K=2$ at $b=0$ fm and $t=3$fm/c.} \label{fig6}
\end{figure}

Another issue raised recently is the polar anisotropy: 
One of reasons is that strong orientation of partons into the collision axis can cause the color
filamentation and the momentum can be further isotropic because of the filamentation effects\cite{mro96}.
The other is that once the momenta of partons are isotropic, 
the hydrodynamic equation of motion can be applicable
to describe the evolution even though the system is not (locally) thermalized\cite{arn04}.
This polar isotropy can be studied also with this cascade program. We first of all can define
the polar anistropy as a function of ploar angle as follows:
Consider the number density into a solid angle $d\Omega$,
\begin{eqnarray}
n(\theta, \phi) &=& {1 \over {N_T}} {{dN}\over{d\Omega}} - {1 \over 2},
\end{eqnarray}
then we can define the normalized density such that
\begin{eqnarray}
\bar n (\theta, \phi) &=&  n (\theta, \phi) / N_T,
\end{eqnarray}
where $N_T$ is the total number of partons.
When the system is isotropic, $\bar n = 1/ 4\pi $.
We integrate over azimuthal angle, which is approximately isotropic, and can obtain the normalized anisotropy
density
$\varepsilon = \bar n (\eta) - 1/2 $ where $\eta = cos \theta$. 
Fig \ref{fig6} shows the polar anisotropy at $t=3$ fm/c.
\\
The figure shows that the system never have polar isotropy even in the central rapidity so that
the condition for the filamentation can be met but that of applicability of hydrodynamics is not.

\section{Jet quenching}
\begin{figure}
\includegraphics{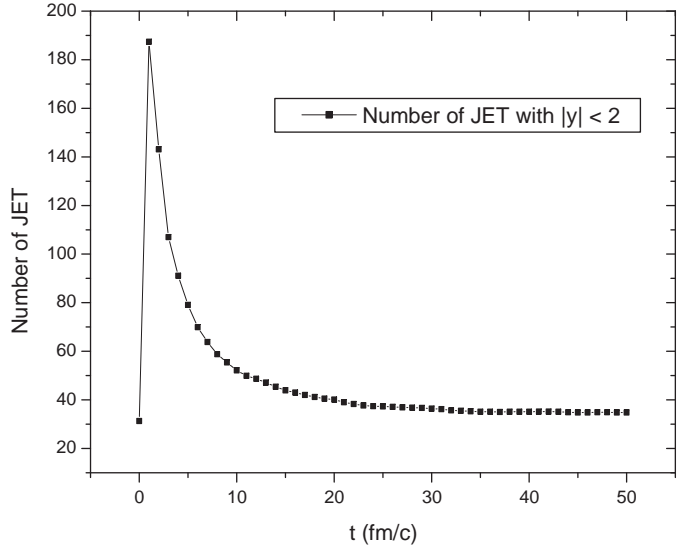}
\caption{The number of jets as a function of time: formation time=0.1 fm, K=2, b=0 fm.} \label{fig7}
\end{figure}

High energy parton can loose its energy by two mechanism\cite{bm86,wg92,bd97,gv03,kw03}.
The first one is gluon radiation and the other is collisional energy loss. 
Study shows that the elastic scattering, 
which is major channel of partonic level, causes far less energy loss than the radiation.
On the other hand, the system formed after relativistic heavy ion collision is rapidly expanding
and those high energy partons of running to the collisional axis do not make any serious
scattering but those of high transversal momentum have a chance to go through 
the expanding medium and lose substantial amount of energy by inelastic scatterings which
could cause the gluon radiation.
The parton cascade code(pcc), which we are using, has the radiation channel, Ed. \ref{gg_ggg},
and the available energy for this channel is high enough, several GeV, so that it is
ideal tool to study the jet quenching. We simulate the given 10 sets of initial phase space distribution
and average over the sets. The initial phase space from CGC shattering mechanism
does not have back-to-back high energy parton. We thus calculate the number of 
high transvesal partons, $p_T > 6$ GeV/c, as a function of time and obtain the attenuation rate.
Fig. \ref{fig7} shows the number of high $p_T$ partons with 0.1 fm/c formation time as a function of time.
\\

We can see that the number of jets increases in early stage of evolution, $t < 1 fm/c $,
and decreases rapidly until 10 fm/c since then. This shows that those high energy partons, which trevel
to  longitudinal (collisional) direction, make collisions among them and produces many more
high $p_T$ jets. These jets make further collisions with medium within 10 fm/c and lose energy.
The total attenuation rate is estimated to about 1/3 or 1/4.

\begin{figure}
\includegraphics{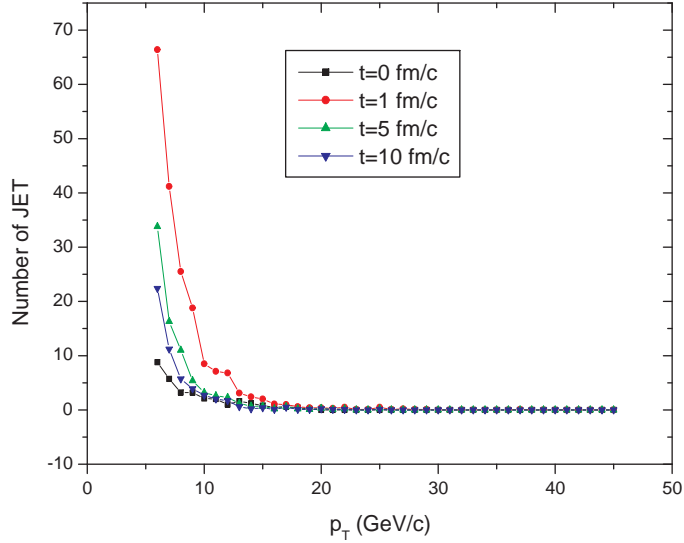}
\caption{$p_T$ distribution of jets.} \label{fig8}
\end{figure}

Fig. \ref{fig8} shows the $p_T$ distribution of jet.
These high energy or high $p_T$ partons eventually fragment and become relatively low
energy or low $p_T$ hadrons. We hope to resolve the theoretical issue on the hadronization from
a parton(s) and apply to this study.
\\

\section{Summary and Conclusions}
We have shown that the elliptic flow can be reproduced from the kinetic theory with reasonable
initial distribution and the polar anisotropy shows that the system is not isotropic enough to apply
a hydrodynamics but the momenta of partons, especially high energy partons, are strongly oriented
to the forward or backward direction which gives the color filamentation mechanism applicable.
We have shown that the attenuation rate of high transverse momenta jets is about 3 or 4.
But we cannot give any conclusive result from this since we expect the jet will fragmetate and hadronize
substantially.\\
We plan in near future to improve the statistics by using more sets of test partons and implement the hadronization
mechanism which is recombination for low energy partons or fragmentation for high energy ones.
It could bring us more detailed understanding on the relativistic heavy ion collisions and
the fundamental signals of quark-gluon plasma.\\

\begin{acknowledgments}
This work was supported by Korea Research Foundation Grant (KRF-2004-015-C00120).
\end{acknowledgments}

\end{document}